\documentclass[aps,prl,twocolumn,groupedaddress]{revtex4-1}

\usepackage{graphicx}
\usepackage{braket}
\usepackage{amsmath,amssymb}
\usepackage{bm}
\usepackage{ulem}
\usepackage{dcolumn}
\usepackage{color}

\newcommand{\nucl}[2]{{}^{#1}\mathrm{#2}}
\newcommand{\CC}{\nucl{12}{C} + \nucl{12}{C}}
\newcommand{\CO}{\nucl{12}{C} + \nucl{16}{O}}

\begin{document}


\title{$\mathbf{{}^{12}{C} + {}^{12}{C}}$ Fusion $\bm{S^*}$-factor from a Full-microscopic Nuclear
Model}  


\author{Yasutaka Taniguchi}
\email[]{taniguchi-y@di.kagawa-nct.ac.jp}
\affiliation{Department of Information Engineering, National Institute of Technology
(KOSEN), Kagawa College, Mitoyo, Kagawa 769-1192, Japan}
\affiliation{Research Center for Nuclear Physics (RCNP), Osaka University, Ibaraki
567-0047, Japan}
\author{Masaaki Kimura}
\email[]{masaaki@nucl.sci.hokudai.ac.jp}
\affiliation{Department of Physics, Hokkaido University, Sapporo 060-0810, Japan} 
\affiliation{Nuclear Reaction Data Centre, Hokkaido University, Sapporo 060-0810, Japan} 
\affiliation{Research Center for Nuclear Physics (RCNP), Osaka University, Ibaraki 567-0047, 
Japan}

\date{\today}

\begin{abstract}
 The $\CC$ fusion reaction plays a vital role in the explosive phenomena of the universe. 
The resonances in the Gamow window rule its reaction rate and products. 
Hence,
 the determination of the resonance parameters by nuclear models is indispensable as the direct 
 measurement is not feasible. Here, for the first time, we report the resonances in the $\CC$ fusion 
 reaction described by a full-microscopic nuclear model. The model plausibly reproduces the measured low-energy  astrophysical $S$-factors and predicts the resonances in the Gamow window. Contradictory to the hindrance model, we conclude that there is no low-energy suppression
 of the $S$-factor.
\end{abstract}


\maketitle

\textit{Introduction.}---
The $\CC$ fusion reaction is a trigger and driving force of the carbon burning in massive
stars~\cite{Rolfs1988,RevModPhys.74.1015} and the X-ray
superburst~\cite{Cumming2001,Strohmayer2002}. Hence, its reaction rate is a key for   
understanding these explosive phenomena. In general, the reaction rate of charged
particles is represented by a product of the exponentially damping factor and the astrophysical
$S$-factor~\cite{Rolfs1988}. While the former represents the Coulomb penetrability, the 
latter carries the nuclear structural information consists of the resonant and non-resonant 
contributions. Because the direct measurement is quite difficult, the $S$-factor of the $\CC$
reaction at low-energy has been extrapolated from the  measurement at higher energy. As a result, it has been
the source of considerable uncertainty despite decades of studies~\cite{Back2014a,Beck2020}.  
An estimate by Caughlan and Fowler (CF88)~\cite{Caughlan1988}, a de facto standard for astrophysics
simulations, assumes constant $S$-factor. On the contrary, Jiang et
al. proposed the hindrance model~\cite{Jiang2007}, which asserts reducing the $S$-factor at
low energy. It is still controversial which one is reasonable.

In addition to the uncertainty in the global behavior, the fine structure of the $S$-factor
originating in the resonant contribution is even less understood. It is well known that low-energy resonances can significantly affect the evolution of astrophysical 
phenomena by increasing the reaction rate in orders of magnitude at specific
temperatures~\cite{Rolfs1988}. For example, Cooper et al. introduced an ad-hoc
resonance at 1.5~MeV inside the Gamow window of X-ray superburst~\cite{Cooper2009}. In nuclear physics, such low-energy resonances  in
the $\CC$ system have long been
discussed~\cite{Almqvist1960,Stokstad1976,Korotky1979,Freer2007}. Recently a couple of direct
measurements have identified the resonances just above the Gamow
window~\cite{Spillane2007,Bucher2015,Tan2020,Fruet2020}. Furthermore, from the indirect measurement,
Tumino et al. reported numerous narrow resonances in the Gamow window~\cite{Tumino2018}. Contrary to
the hindrance model, they proposed the reaction rate enhanced at low energy, albeit its absolute 
magnitude is still under debate~\cite{Mukhamedzhanov2019}.  

Thus, the resonances in the $\CC$ system are of particular interest. Since they are unlikely be
measurable by the direct reactions, the study by nuclear models is indispensable. However, the
description of deep sub-barrier resonances is challenging  and demanding. The primary reaction
channels,  
$^{12}\mathrm{C}(^{12}\mathrm{C},\alpha){}^{20}\mathrm{Ne}$ and 
$^{12}\mathrm{C}(^{12}\mathrm{C},p){}^{23}\mathrm{Na}$, involve the rearrangement of many nucleons
and the strong channel coupling. To mimic such complex reaction dynamics, nuclear models usually
adopt phenomenological potentials~\cite{Gasques2007, Esbensen2011, Diaz-Torres2018, Chien2018}, undermining their predictability at low energy. In principle, full-microscopic nuclear models
without phenomenologically adjustable parameters can overcome this problem~\cite{Godbey2019}, but
the resonant contributions have never been taken into account. Here, for the first time, we report a  
plausible description of the resonances by a full-microscopic nuclear model and provide an
evaluation of the $S$-factor.   

\textit{Nuclear model and resonance parameters.}---
To describe the low-energy resonances, we employ  antisymmetrized molecular dynamics
(AMD)~\cite{Kanada-Enyo2012}, which handles the channel coupling by configuration mixing. Using the Gogny D1S nuclear density functional \cite{Berger1991}, it
accurately describes the low-energy resonances of astrophysical interests such as $\CO$ and  $\alpha
+ \nucl{28}{Si}$ and their compound
systems~\cite{Kimura2004,Taniguchi2009,Taniguchi2020,Kimura2020}. The model wave function of AMD is
a parity-projected  Slater determinant of the nucleon wave packets~\cite{Kimura2004a},     
\begin{align}
  \Phi^\pi &= \frac{1+\pi P_r}{2}{\mathcal{A}}\{\varphi_1\cdots\varphi_A\}, \quad(\pi=\pm),\\
 \varphi_i&=\prod_{\sigma=x,y,z}
 e^{-\nu_\sigma\left(r_\sigma - Z_{i\sigma}\right)^2}
 \left(a_i\ket{\uparrow}+b_i\ket{\downarrow}\right)\eta_i,
\end{align}
where each wave packet $\varphi_i$ has the parameters; Gaussian centroid $\bm {Z}_i$, 
width $\bm{\nu}$ and spin $a_i$ and $b_i$. The isospin $\eta_i$ is fixed to proton or neutron.
Using the constraint on the inter-nuclear distance~\cite{Taniguchi2004}, all parameters are determined for
each channel and inter-nuclear distance by the energy variation. In this study, we have calculated
the $\CC$ and $\alpha+\nucl{20}{Ne}$ channels by using this method.   
\begin{figure}[tbp]
 \begin{center}
\begin{tabular}{ccc}
 \includegraphics[width=0.15\textwidth]{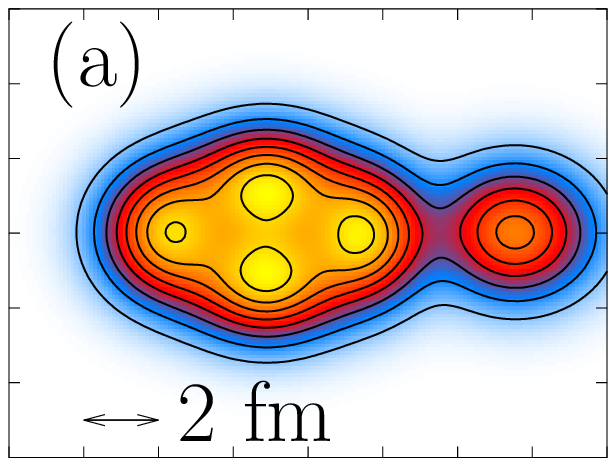}&
 \includegraphics[width=0.15\textwidth]{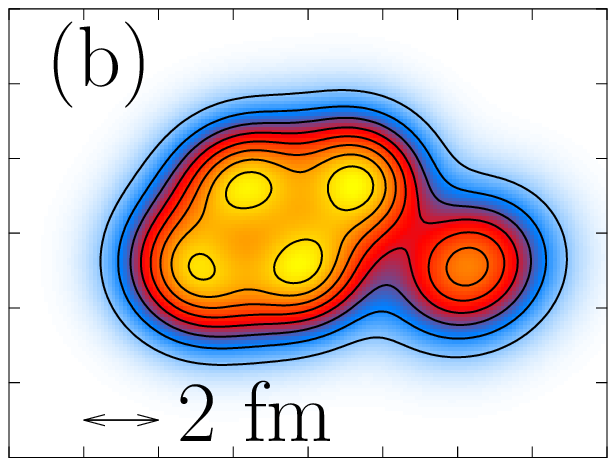}&
 \includegraphics[width=0.15\textwidth]{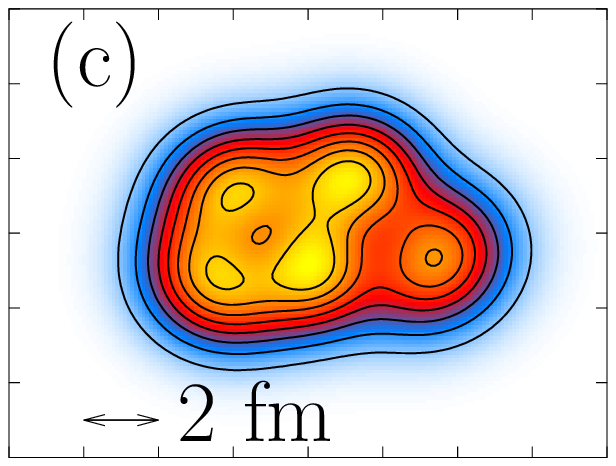}\\
 \includegraphics[width=0.15\textwidth]{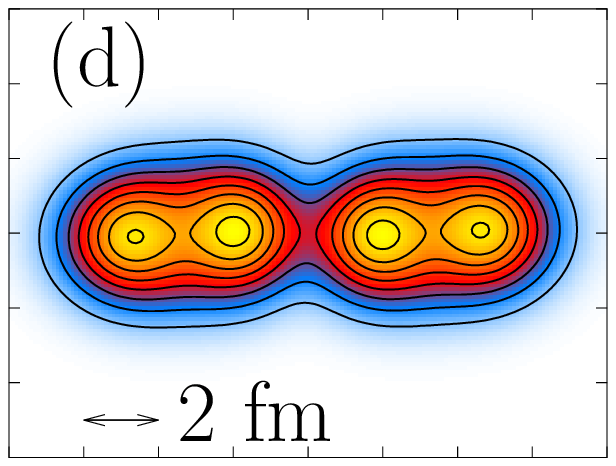}&
 \includegraphics[width=0.15\textwidth]{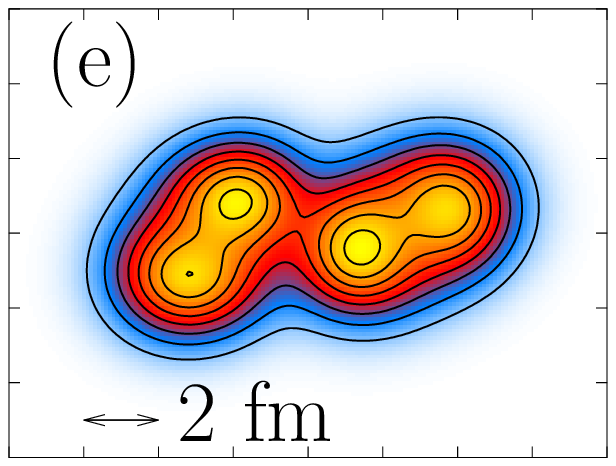}&
 \includegraphics[width=0.15\textwidth]{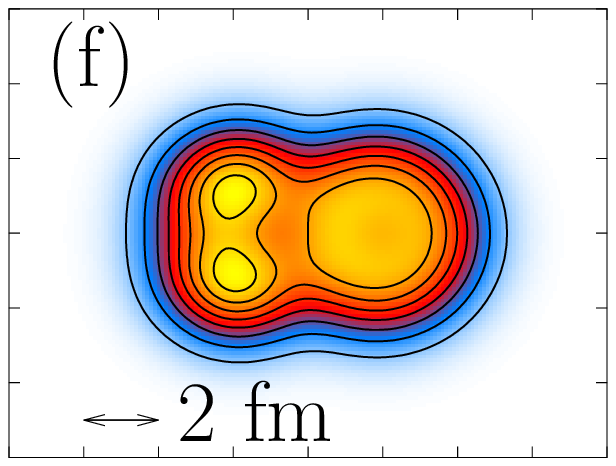}\\
\end{tabular}  
\caption{
  The wave functions for the (a)--(c) $\alpha + \nucl{20}{Ne}$, and (d)--(f) $\CC$ channels obtained by the energy variation. 
  Inter-nuclear distance is 6.5, 5.0, and 3.5~fm from left to right.
  } \label{density}
 \end{center}
\end{figure}
Figure~\ref{density} shows several channel wave functions which have different
inter-nuclear distances and orientations of nuclei. Note that the rotation and polarization
of nuclei depending on the inter-nuclear distance are described naturally. As explained
later, these are essential in describing the low-energy resonances. In addition, we also include
the wave functions of the compound nucleus $^{24}{\rm Mg}$~\cite{Kimura2012}. All these channel wave
functions  ($\Phi^\pi_i$) are superposed to describe the resonances,   
\begin{align}
 \Psi_{M}^{J\pi} = \sum_{iK}c_{iK}P^{J}_{MK}\Phi^\pi_i,
\end{align}
where $P^{J}_{MK}$ denotes the angular momentum projector. The resonance energy and the coefficients 
$c_{iK}$ are determined by the diagonalization of the Hamiltonian. The superposition of various
channel wave functions significantly improves the model accuracy. First, it describes the rotational
excitation of nuclei during the reaction process. Therefore, the model can explain the decays to
the excited states such as $\alpha+{}^{20}\mathrm{Ne}(2^+_1)$. It also accounts for the dynamical change
of the Coulomb barrier height due to the nuclear deformation~\cite{Hagino2012}. Second, the channel
coupling is described by a full-microscopic Hamiltonian in a unified manner. Hence, the model is
free from adjustable parameters. These improvements realize an accurate description of the deep
sub-barrier resonances.  

To evaluate  the decay widths, we calculate the reduced width amplitude (RWA)~\cite{Balashov1959,Chiba2017} which
is the overlap between the decay channel and resonance wave functions,   
\begin{align}
 y_l^{A_1 + A_2}(a) = \sqrt{\tbinom{24}{A_1}}
 \Braket{\frac{\delta(r-a)}{a^2}[\Phi_{A_1}\Phi_{A_2}Y_{l}(\hat r)]^J_{M}|
 \Psi^{J\pi}_M},\label{eq:rwa1}
\end{align}
where the ket is the resonance wave function, while the bra is the channel wave function decaying
into two nuclei with masses $A_1$ and $A_2$ separated by the distance $a$ with the orbital angular
momentum $l$.  We consider four decay channels, $p+\nucl{23}{Na} (3/2^+_1)$, 
$p+\nucl{23}{Na^*}(5/2^+_1)$, $\alpha+\nucl{20}{Ne}(0^+_1)$ and $\alpha+\nucl{20}{Ne^*}(2^+_1)$ which 
are denoted by $p_0$, $p_1$, $\alpha_0$ and $\alpha_1$, respectively. The wave functions of
$\alpha$, $\nucl{20}{Ne}$, and $\nucl{23}{Na}$ are also calculated by AMD. For the later use, the
RWAs are given as the ratio to the Wigner limit, 
\begin{equation}
 \theta^2_{A_1 + A_2} (a) = \frac{a}{3} \left|a y_l^{A_1 + A_2} (a)\right|^2,
\end{equation}
where the channel radii for the $\CC$, $\alpha_{0,1}$, and $p_{0,1}$ channels are chosen as 6, 6 and
4~fm so that RWAs are smoothly connected to the Coulomb wave function.  

 \begin{table*}[!ht]
  \caption{The calculated resonance energies in MeV, isoscalar transition matrix elements in 
  Weisskopf unit, and RWAs of the $0^+$ and $2^+$ resonances. The RWAs are given as the ratio
  to the Wigner limit (see text) and multiplied by a factor of a hundred.}\label{tab:res}   
 \begin{ruledtabular}
  \begin{tabular}{ccccccccccccccc}
   & & &$\theta_\mathrm{C}^2\times 10^2$
   &\multicolumn{2}{c}{$\theta_{\alpha_0}^2\times 10^2$}
   &\multicolumn{3}{c}{$\theta_{\alpha_1}^2 \times 10^2$} 
   &\multicolumn{3}{c}{$\theta_{p_0}^2\times 10^2$} 
   &\multicolumn{3}{c}{$\theta_{p_1}^2 \times 10^2$}\\
    \cline{5-6}\cline{7-9}\cline{10-12}\cline{13-15}
   $J^\pi$& $E_\mathrm{R}$& $M_{\rm IS}$ & $l = J$ &
   $l = 0$ & $2$ & $l=0$ & $2$ & $4$ &
   $l = 0$ & $2$ & $4$ & $l=0$ & $2$ & $4$ \\\hline
   $2^+$ & 0.93  & 1.56 & $1.4$ &
   ---     & $3.5$  & $0.061$  & $1.7$  & $6.7  $  &
   $ 0.47$ & $0.081$  & $0.030$ & $0.20$  & $0.15$  & $0.083$\\
   $0^+$ & 0.94  & 0.59 & $7.3$ &
   $ 0.20$ &  ---    &  ---   & $ 7.1$  &  ---    &
   ---     & $0.10$  &  ---    &   ---   & $0.69$  &  ---    \\ 
   $2^+$ & 1.50  & 1.04 & $2.9$ &
   ---     & $1.1$  & $ 4.0 $ & $0.90$  & $ 0.51$  &
   $ 0.16$ & $0.22$  & $0.001$  & $0.012$ & $0.098$ & $0.005$ \\
   $2^+$ & 2.18 & 0.51 & $3.4$ &
   ---     & $1.0$  & $ 1.0$  & $0.19$  & $ 3.4$  &
   $ 3.3$ & $0.12$  & $0.010$  & $0.70$  & $0.11$  & $0.23$  \\
   $0^+$ & 3.02 & 1.05 & $11$ &
   $ 0.26$ &  ---    &  ---   & $0.57$  &  ---    &
   ---     & $0.99$  &  ---    &   ---   & $0.43$  &  ---    \\ 
   $2^+$ & 3.56 & 0.23 & $1.2$ &
   ---     & $0.038$  & $0.056$ & $0.006$ & $0.040$ &
   $ 0.66$ & $0.86$  & $0.001$  & $0.029$ & $0.79$  & $0.041$ \\
   $2^+$ & 3.73 & 0.41 & $8.3$ &
   ---     & $0.10$  & $0.066$ & $0.10$  & $ 0.88$ &
   $ 0.24$ & $0.72$  & $0.028$  & $0.043$ & $0.67$  & $0.089$ 
  \end{tabular}
 \end{ruledtabular}
\end{table*}

 
Table~\ref{tab:res} lists the calculated resonance parameters in the energy range of interest
(resonance energy $E_\mathrm{R}<4$~MeV). Note that the calculation yields many deep sub-barrier resonances. In
particular, it is remarkable that some of them plausibly coincide with the observed higher-energy peaks of the
$S$-factor at higher energies. For example, the $J^\pi = 2^+$ resonances obtained at 2.18 and
3.73~MeV are the candidates for the 2.14 and 3.8~MeV resonances identified by the direct 
measurements~\cite{Kettner1980a,Spillane2007,Fruet2020}. Furthermore, the calculation predicts three resonances
within the Gamow window at 0.93, 0.94, and 1.50~MeV. This result confirms the
long-standing conjecture about the existence of the low-energy resonances in the $\CC$
system~\cite{Almqvist1960,Stokstad1976,Korotky1979} and is in accordance with the indirect measurement~\cite{Tumino2018} which
reported many resonances.

The RWAs of the resonances demonstrate the importance of the microscopic treatment of channel
coupling and rotational excitation. The $0^+$ resonance at 0.94~MeV has a large RWA in the $\alpha_1$
channel showing the rotational excitation of $^{20}{\rm Ne}$. Moreover, the $2^+$ resonance at 2.18
MeV has sizable RWAs in the $\alpha_0$, $\alpha_1$, and $p_0$ channels and in the entrance
$\CC$ channel. Therefore, it should affect the reaction rates in both the $\alpha$ and $p$
channels. The importance of the channel coupling can also be confirmed differently. If we
perform a single-channel calculation, for example, with only the $\CC$ channel, we do not obtain any  
resonance in the Gamow window. It demonstrates the necessity of the full-microscopic
calculation for quantitative discussion of the sub-barrier resonances.  

\begin{figure}[htbp]
 \begin{center}
  \includegraphics[width=0.45\textwidth]{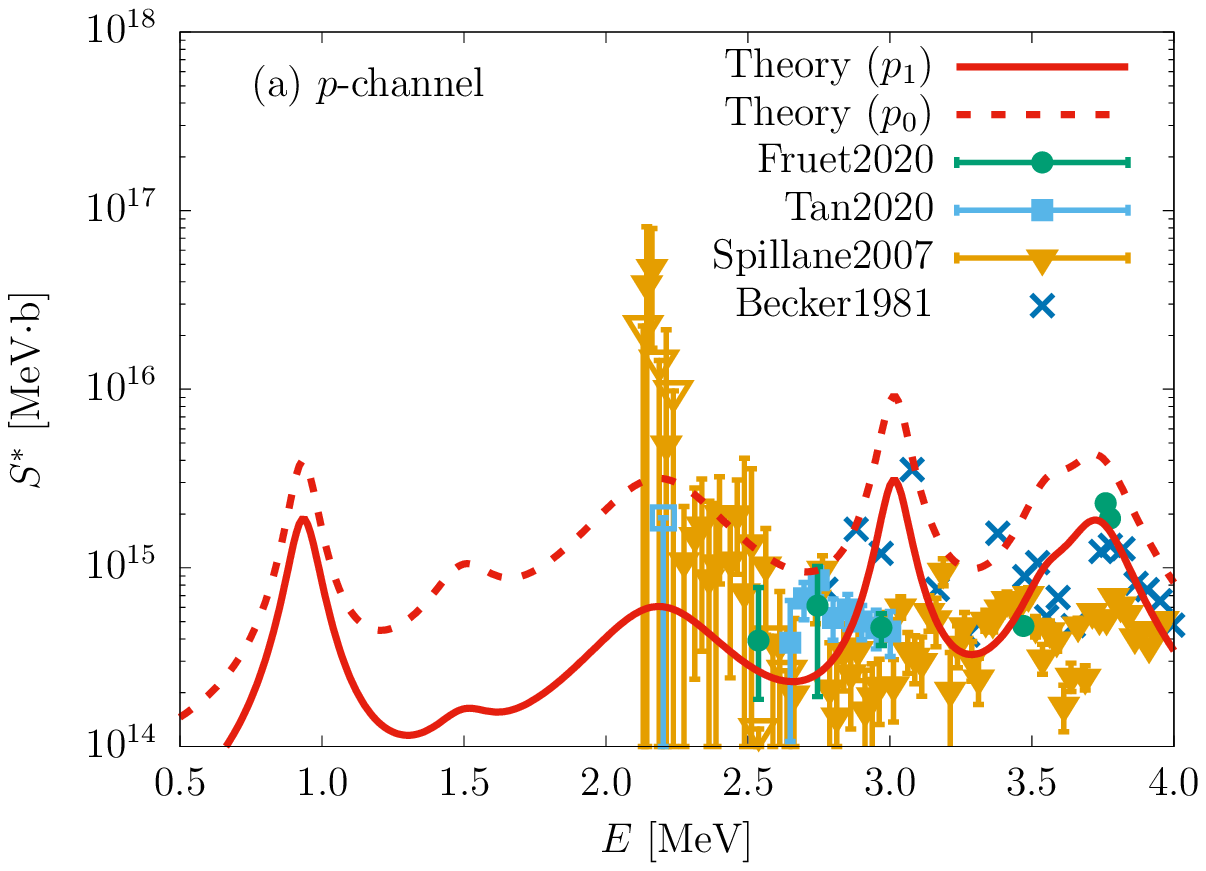}\\
  \includegraphics[width=0.45\textwidth]{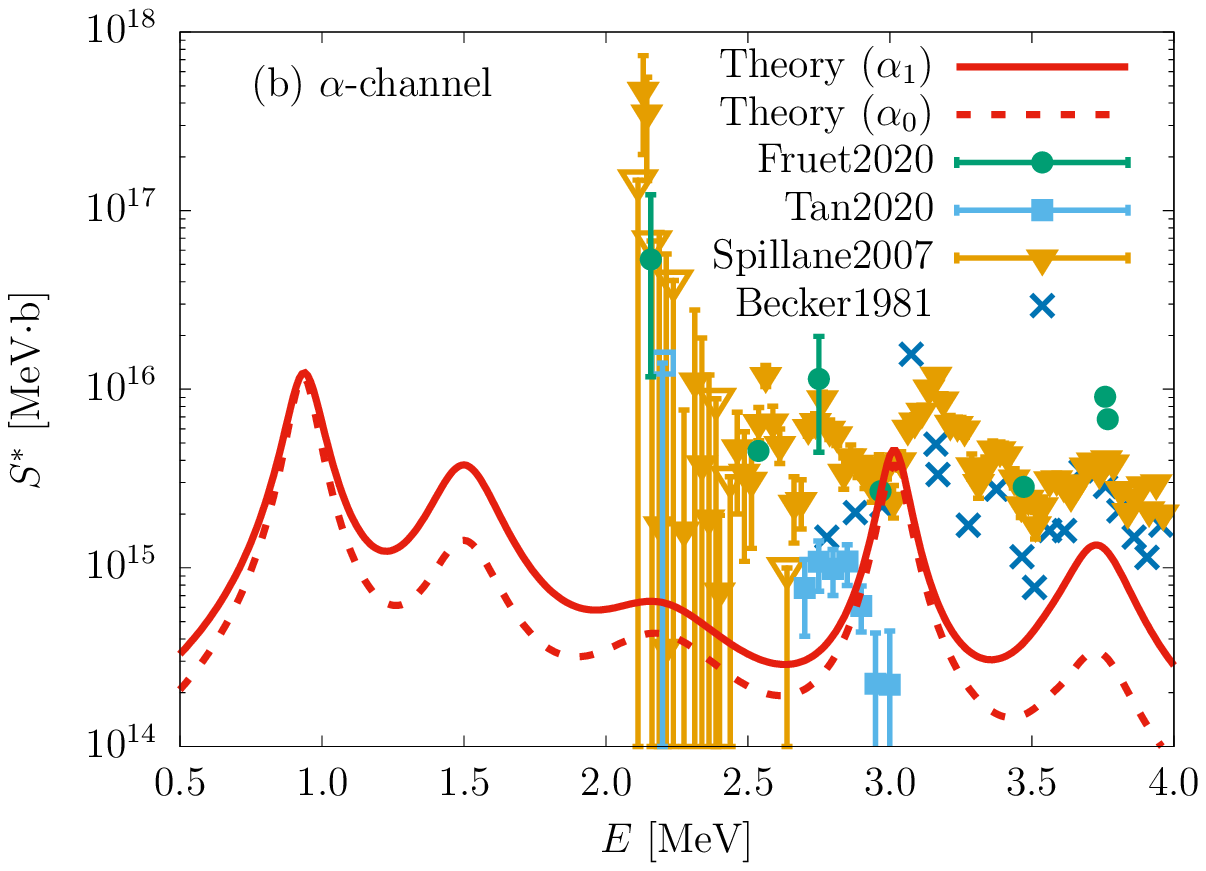}\\
  \includegraphics[width=0.45\textwidth]{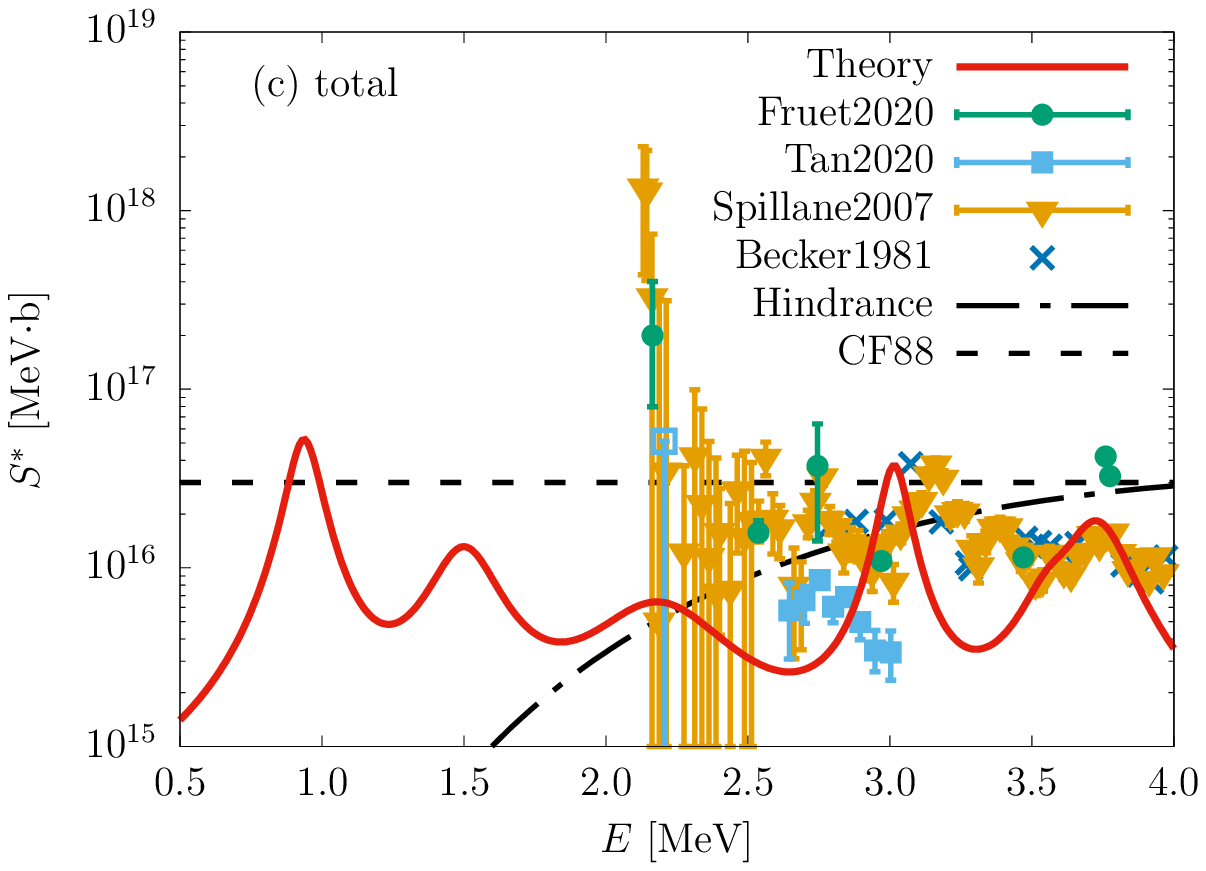}
  \caption{Calculated and observed~\cite{Becker1981,Fruet2020,Tan2020,Spillane2007} modified
  astrophysical $S^*$-factors in the (a) $p_1$ and (b) $\alpha_1$ channels. The calculated
  $S^*$-factors in the $p_0$ and $\alpha_0$ channels are   also shown for the comparison. Panel (c)
  compares the calculated and observed total $S^*$-factors with the
  evaluations~\cite{Jiang2007,Caughlan1988}. Open symbols indicate upper limits.} \label{S_factor}   
 \end{center}
\end{figure}

\textit{Astrophysical S-factor.}---
From the resonance parameters, we evaluate the modified astrophysical $S$-factors ($S^\ast$-factors) \cite{Patterson1969}. According to the $R$-matrix
theory~\cite{Lane1958,Descouvemont2010}, the partial width for the $A_1 + A_2$ decay is given as,  
\begin{equation}
 \Gamma_{A_1 + A_2} = \frac{2 ka}{F_l (k a)^2 + G_l (k a)^2} \frac{3\hbar^2}{2\mu a^2} \theta_{A_1 + A_2}^2,
\end{equation}
where the wave number $k$ is related with the decay $Q$-value as $Q = \hbar^2 k^2 / 2\mu$ with the
reduced mass $\mu$. $F_l$ and $G_l$ are the regular and irregular Coulomb wave functions, respectively.
The partial cross section at the center-of-mass energy $E$ is given as the Breit-Wigner form~\cite{Lane1958},
\begin{equation}
 \sigma (E) = \frac{\pi\hbar^2(2J+1)}{2\mu E}
  \frac{\Gamma_\mathrm{I} \Gamma_\mathrm{A_1+A_2}}{\left(E - E_\mathrm{R}\right)^2 + \Gamma^2/4},
\end{equation}
where $J$ and $E_\mathrm{R}$ denote the spin and energy of a resonance and $\Gamma_\mathrm{I}$ is the
partial width in the entrance ($\CC$) channel. The total width $\Gamma$ is
estimated in the same manner as in the latest experimental analysis~\cite{Tan2020,Fruet2020}, 
i.e., it is estimated from the $p_1$ and $\alpha_1$ widths and the linear extrapolation
of the observed branching ratio~\cite{Becker1981}. The $S^\ast$-factor is calculated
as,  
\begin{equation}
 S^\ast(E) = E \sigma(E) \exp(2\pi \eta + 0.46~\mathrm{MeV}^{-1} E),
\end{equation}
with  the Sommerfeld parameter $\eta =  36/137\sqrt{\mu c^2/2 E_\text{R}}$ \cite{Patterson1969}.

Figure~\ref{S_factor}(a) compares the calculated and observed $S^*$-factors in the $p_1$ channel.
Above the Gamow window ($E \gtrsim 2.0$~MeV), the calculated $S^*$-factor has the contributions from
four resonances (2.18, 3.02, 3.56, and 3.73~MeV) and its magnitude is the order of
$10^{14}$--$10^{15}~\mathrm{MeV\cdot b}$, which plausibly agrees with the
experiments~\cite{Spillane2007,Tan2020,Fruet2020,Becker1981,Kettner1980a}.
The peak positions at 2.2 and 3.8~MeV coincide with the
direct $\CC$ fusion experiments~\cite{Kettner1980a,Spillane2007,Fruet2020}. 
The calculation predicts that the resonant contributions are also present inside the Gamow
window. In particular, the resonances at 0.94 and 0.95~MeV form a prominent peak. For the
comparison, the calculated $S^*$-factor in the $p_0$ channel is also shown in Fig.~\ref{S_factor}(a). It has a similar shape, but the magnitude is much larger than the $p_1$ channel.

Because of the strong coupling between $p$ and $\alpha$ channels, the same resonant peaks appear in
the $\alpha_1$ channel, as shown in Fig.~\ref{S_factor}(b). 
The peaks at 2.2 and 3.8~MeV exist in
the $\alpha_1$ channel in the calculation and observed
data~\cite{Kettner1980a,Becker1981,Spillane2007}.
The calculation yields more prominent peak at 3.0~MeV than that in the $p_1$ channel. This
peak nicely coincides with the peak approximately at 3.2~MeV observed in
Refs.~\cite{Kettner1980a,Becker1981,Spillane2007}. 
It is impressive that the calculation predicts pronounced $S^*$-factor inside the Gamow window even
stronger than above the window owing to the contributions from the resonances at  0.93 and 0.94~MeV. 

Finally, Fig.~\ref{S_factor}(c) shows the total $S^\ast$-factor as the sum of the
$S^*$-factors in all $\alpha$ and $p$ channels. Here, the total $S^*$-factors in the $\alpha$ and $p$
channels are again estimated in the same manner with Refs.~\cite{Tan2020,Fruet2020}.
The figure also shows the estimations by CF88~\cite{Caughlan1988} and the hindrance
model~\cite{Jiang2007} in addition to the data from
Refs.~\cite{Becker1981,Spillane2007,Tan2020,Fruet2020}. Similar to the $S^*$-factors in the $p_1$
and $\alpha_1$ channels, the calculation plausibly describes the observed global behavior. Furthermore, as
we have already repeated, the calculation predicts resonance contributions at very
low energy. Consequently, the calculated $S^*$-factor has prominent    peaks in the Gamow window and
does not show the low-energy hindrance. Hence, the low-energy enhancement of the $S$-factor, for
example, that postulated by Cooper et  al. \cite{Cooper2009}, is conceivable enough.

\textit{A novel probe for the resonances.}---
Finally, we propose the isoscalar transitions as a novel experimental probe to identify the resonances
predicted in this study. The transition matrix element from the ground state to a resonance is 
defined as,
\begin{align}
 M_\mathrm{IS} = \braket{J^\pi,J_z=0|\mathcal{M}_\mathrm{IS}|\mathrm{^{24}{Mg}(g.s.)}},
\end{align}
where the bra is the resonance wave function with the spin-parity $J^\pi$, while the ket is the
ground state of $^{24}{\rm Mg}$.  $\mathcal{M}_\mathrm{IS}$ is either of the isoscalar monopole or
quadrupole operators depending on the spin-parity of the resonance, 
\begin{align}
 \mathcal{M}_\mathrm{IS} = \left\{
 \begin{array}{l}
  \sum_{i=1}^A r_i^2, \quad J^\pi=0^+\\
  \sum_{i=1}^A r_i^2 Y_{20}(\hat r_i), \quad J^\pi=2^+.
 \end{array}
 \right.
\end{align}

In recent years, it has been extensively discussed that the isoscalar transitions from the ground
state to the cluster resonances are considerably
enhanced~\cite{Kawabata2007,Kanada-Enyo2007,Yamada2008,Chiba2016}. 
As listed in Table~\ref{tab:res}, the present calculation predicts that  all the resonances have
large transition strengths comparable with  the Weisskopf  unit. Therefore, the reactions that
induce isoscalar transitions should strongly 
populate these resonances. The most   promising and feasible reaction  may be the $\alpha$-inelastic
scattering,  $^{24}{\rm Mg}(\alpha,\alpha')^{24}{\rm Mg}^*$. Note that this reaction bypasses the
Coulomb barrier, and hence, can easily access the deep sub-barrier resonances. It is encouraging
that a couple of experiments have already been
conducted~\cite{Youngblood2009,Kawabata2013,Gupta2015,Adsley2021}, and  several candidates of the
resonances are observed in the energy range of our interest, although the spin-parity and decay 
branches have not been confirmed firmly. More detailed analysis and the comparison with the
calculations must uncover the resonances in the Gamow window.

\textit{Summary.}---
We have investigated the low-energy resonances in the $\CC$ fusion reaction, which is of
astrophysical interest. The AMD, a microscopic nuclear model, firstly realized a quantitative
description of the low-energy resonances by handling the channel coupling and nuclear rotation
without adjustable parameters. It successfully described observed resonances and behavior of the
$S^*$-factors above the Gamow window. Furthermore, it predicted several resonances inside the Gamow
window, which create low-energy $S^*$-factor peaks. Consequently, the calculation suggested no
low-energy suppression of the $S^*$-factor supporting the result of the indirect 
measurement~\cite{Tumino2018}. Finally, we propose the 
$^{24}{\rm Mg}(\alpha,\alpha')^{24}{\rm Mg}^*$ reaction experiment as the most promising  bypass to
access the deep sub-barrier resonances.   

\begin{acknowledgments}
 This work was supported by JSPS KAKENHI Grant Number 19K03859, the collaborative research program
 at Hokkaido University, and the COREnet program in RCNP, Osaka University. 
 Numerical calculations were performed using Oakforest-PACS at the CCS, University of Tsukuba.
\end{acknowledgments}

\bibliography{CC}

\end{document}